# Thickness-dependent anisotropic Gilbert damping in heterostructures of ferromagnets and two-dimensional ferroelectric bismuth monolayer

Shi-Bo Zhao[1], Xiang-Fan Huang[1], Ze-quan Wang[1], Ruqian Wu[2], and Yusheng Hou[1,*]

[1] Guangdong Provincial Key Laboratory of Magnetoelectric Physics and Devices, Center for Neutron Science and Technology, School of Physics, Sun Yat-Sen University, Guangzhou, 510275, China

[2] Department of Physics and Astronomy, University of California, Irvine, CA 92697-4575, USA

**Abstract**

The Gilbert damping parameter, which describes magnetization dynamics, is crucial for the performance of modern spintronic devices, affecting factors such as the switching speed and critical current density of magnetoresistive random access memory. Thus, the ability to engineer it on demand is pivotal for developing novel spintronic applications. In this work, we systematically examine the Gilbert damping parameter of Fe films in contact with a black phosphorus-like bismuth monolayer using first-principles calculations. In these Bi/Fe heterostructures, we obtain a significantly enhanced Gilbert damping owing to strong interfacial spin-orbit couplings (SOCs). Interestingly, we find non-monotonic thickness-dependent Gilbert damping anisotropy and attribute that to the competition between the interfacial SOC and the intrinsically anisotropic SOC of Fe films. We further demonstrate that these SOC effects lead to anisotropic band structures, which are responsible for the anisotropic Gilbert damping. Our work provides a deep understanding of the anisotropic Gilbert damping and opens avenues for exploring it in ferromagnetic heterostructures.

[*] Corresponding authors: houysh@mail.sysu.edu.cn

## I. INTRODUCTION

As a dimensionless phenomenological parameter in the Landau-Lifshitz-Gilbert (LLG) equation [1], the Gilbert damping parameter describes the energy dissipation rate of the magnetization dynamics in ferromagnets and has aroused great interest in the field of spintronics [2-5]. Gilbert damping determines the performance of various spintronic devices, affecting aspects like the response speed and signal-to-noise ratio in hard drives [6,7] as well as the speed of the current-induced magnetization switching and critical current density in certain magnetic devices [8-11]. Thus, tailoring Gilbert damping to meet specific requirements is crucial for developing versatile spintronic devices. Indeed, one can engineer Gilbert damping by tuning the strength of spin-orbit coupling (SOC) $\xi$, the density of states (DOS) $n(E_F)$ at the Fermi level and the momentum scattering time $\tau$ of electrons, all of which are confirmed theoretically and experimentally [12-25]. Additionally, several methods are put forward to tune Gilbert damping parameters, such as interface engineering [26], doping [27] and size effect of nanomagnets [28]. However, their practical application prospects are limited by the fact that they often require demanding working conditions [16,29-32].

With the rapid development of spintronics, more and more endeavors have been devoted to exploring the physics of Gilbert damping anisotropy. Over the past few decades, although the Gilbert damping parameter has been generally assumed to be an isotropic scalar, some theoretical studies have revealed its anisotropic nature and suggested expressing it as a tensor [15,33-36]. Implicitly, the Gilbert damping parameter has a dependence on the instantaneous direction of magnetizations (orientational anisotropy) and on the direction of rotation of magnetizations (rotational anisotropy). Since Gilbert damping associated with electron scattering is believed to mainly consist of intraband and interband relaxations, anisotropic electronic structures can naturally lead to anisotropic Gilbert damping parameters [15]. Although some previous experimental studies attempted to demonstrate the Gilbert damping anisotropy, they all lacked convincing evidence [37-39].

In recent years, a few of intriguing studies have obtained anisotropic Gilbert damping in systems such as Fe/GaAs(001) and Fe/GeTe interfaces as well as in CoFe,

Co-Fe-B(001) and Co-Fe-Al thin films [40-45]. However, beyond considering the effect of SOC and anisotropic DOS, there is a lack of understanding about the microscopic mechanism behind these observations. It is worth noting that the magnitude of anisotropic Gilbert damping in Fe/GaAs can be tuned by changing the thickness of the Fe film. This clearly suggests that the interface effect in this system is important [46]. Besides, it has been demonstrated that through the proximity effects in heterostructures, a material can inherit properties of its neighbors, such as magnetism, superconductivity and strong SOC [47]. Hence, it is conceivable that tunable anisotropic Gilbert damping could also be achieved through the construction of magnetic heterostructures [48]. To this end, it is essential to explore the mechanisms of anisotropic Gilbert damping for effectively regulating it in spintronic devices.

In this work, utilizing first-principles calculations, we systematically study the effect of $\alpha$-phase Bi monolayer (ML), a two-dimensional single-element ferroelectric [49], on the Gilbert damping of ferromagnetic Fe films. Here, Bi ML is chosen for its strong SOC crucial for exploring the Gilbert damping anisotropy via proximity effect, low dimensionality suitable for designing ultrathin devices and lattice constants matching well with those of bcc Fe. When Bi ML and Fe films are integrated into a heterostructure (denoted as Bi/Fe), we find the following interesting phenomena: (i) Compared with a 1ML-Fe, its Gilbert damping is enhanced by about two orders of magnitude; (ii) Its Gilbert damping decreases as the thickness of its Fe films increases; (iii) It has thickness-dependent anisotropic Gilbert damping. We demonstrate that these phenomena are mainly due to the competition between the interfacial SOC and the intrinsic anisotropic SOC of Fe films. Additionally, we show that these SOC effects manifest as anisotropic band structures, which underpin the anisotropic Gilbert damping. Our work provides deep insights into the mechanisms of anisotropic Gilbert dampings and opens up opportunities to realize such effects in ferromagnetic heterostructures.

## II. COMPUTATIONAL DETAILS AND METHODS

Our first-principles calculations based on the density functional theory (DFT) are performed with the Vienna *ab initio* Simulation Package (VASP) at the level of the generalized gradient approximation [50-53]. We treat Bi6*s*6*p* and Fe3*d*4*s* as valence electrons and employ the projector-augmented-wave pseudopotentials to describe core-valence interactions [54]. The energy cutoff of plane-wave expansions is set to 350 eV. In Bi/Fe heterostructures, the *α*-phase Bi ML is simulated by the black phosphorous-like structure (BP-Bi), and Fe films are simulated by a body-centered cubic structure with different thicknesses. To avoid the spurious interactions between adjacent slabs along the normal axis, a vacuum space of 15 Å is added. Positions of all atoms are fully relaxed until the magnitude of the force on each atom is less than 0.01 eV/Å, and a 15×15×1 Γ-centered *k*-point grid is utilized in these procedures. We use the semi-empirical DFT-D3 method to describe the van der Waals interactions in Bi/Fe heterostructures [55].

Based on the scattering theory, Gilbert damping parameters can be determined by first-principles calculation via the linear response formalism [56-58]. The applicability and reliability of this method have been established in previous first-principles studies [19,40,45,58-60]. By extending the torque method for the determination of magnetocrystalline anisotropy, the Gilbert damping parameter can be calculated from the orientation dependence of the Hamiltonian according to the following equation [20]:

$$\alpha_{\mu\nu} = -\frac{\pi\hbar\gamma}{M_S}\sum_{ij} \left\langle \psi_i \left| \frac{\partial \mathbf{H}}{\partial u_\mu} \right| \psi_j \right\rangle \left\langle \psi_j \left| \frac{\partial \mathbf{H}}{\partial u_\nu} \right| \psi_i \right\rangle \times \delta(E_F - E_i)\delta(E_F - E_j) \quad (1)$$

Here, $M_S$ is the saturation magnetization, $\gamma$ is the gyromagnetic ratio, $E_F$ is the Fermi energy, $E_i$ is the energy of band *i*, $u_\mu$ is the deviation of a normalized magnetic moment away from its equilibrium, and the summation is running over all states. To ensure the numerical convergence of Gilbert damping parameters, we increase the *k*-point grid to 65×65×1. Because our DFT calculations show Bi/Fe heterostructures have an in-plane magnetic easy axis [Fig. S1 in Supplemental Material] and the in-plane magnetization dynamics are of importance for the performance of many modern magnetic devices, we focus on studying the dependence of Gilbert dampings on the in-plane magnetization through changing the in-plane magnetization orientation of Fe films. Note that the *a* axis is set as the reference axis for the in-plane magnetization orientation in Bi/Fe heterostructures [see the inset in Fig. 1(b)].

## III. RESULTS AND DISCUSSIONS

Recently, it has been shown that BP-Bi ML is a two-dimensional single-element ferroelectric material [49]. Unlike the homogenous orbital configuration seen in the black phosphorous, Bi atoms in BP-Bi ML exhibit a weak hybridization between 6*s* and 6*p* orbitals. This leads to an inversion-symmetry-broken structure with the orthogonal *Pm* space group and, consequently, the two-dimensional ferroelectricity [49]. As shown in Fig. 1(a), BP-Bi ML has a small buckling (i.e., $\Delta h \neq 0$) and consists of diatomic layers, each of which forms a zigzag chain. For Bi/Fe heterostructures, we add Fe films with different thicknesses on the Bi ML. For the convenience of discussions below, we denote the heterostructure of Fe ML and BP-Bi as Bi/1ML-Fe, that of Fe bilayer and BP-Bi as Bi/2ML-Fe, and so on. It is worth noting that Fe ML structure (denoted as 1ML-Fe) is obtained by directly removing Bi from Bi/1ML-Fe. Fig. 1(b) illustrates the relaxed crystal structure of Bi/1ML-Fe, where one may see that Fe atoms at different positions on BP-Bi ML have different heights. As expected, Fe atoms over the hollow site relax more toward BP-Bi ML, while Fe atoms at the top site of Bi atoms are slightly higher in position. Additionally, the small buckling of BP-Bi ML induced by the weak hybridization between 6*s* and 6*p* orbitals vanishes in Bi/1ML-Fe [i.e., $\Delta h = 0$ in Fig. 1(b)]. This means that the in-plane ferroelectric polarization disappears in Bi/1ML-Fe. Fig. S2 in Supplemental Material (SM) shows the relaxed structures of Bi/Fe with the thickness of Fe films varying from two to eight MLs. We see that the small buckling of BP-Bi ML disappears in all Bi/Fe heterostructures. To describe the in-plane magnetization orientation in Bi/Fe heterostructures, we define an angle $\theta_H$ with $\theta_H = 0°$ corresponding to the magnetization along the *a* axis [see the inset in Fig. 1(b)]. It is worth noting that the symmetry of all Bi/Fe heterostructures is the space group *Pm* which has a mirror symmetry with respect to the *a* axis [Fig. 1(b)].

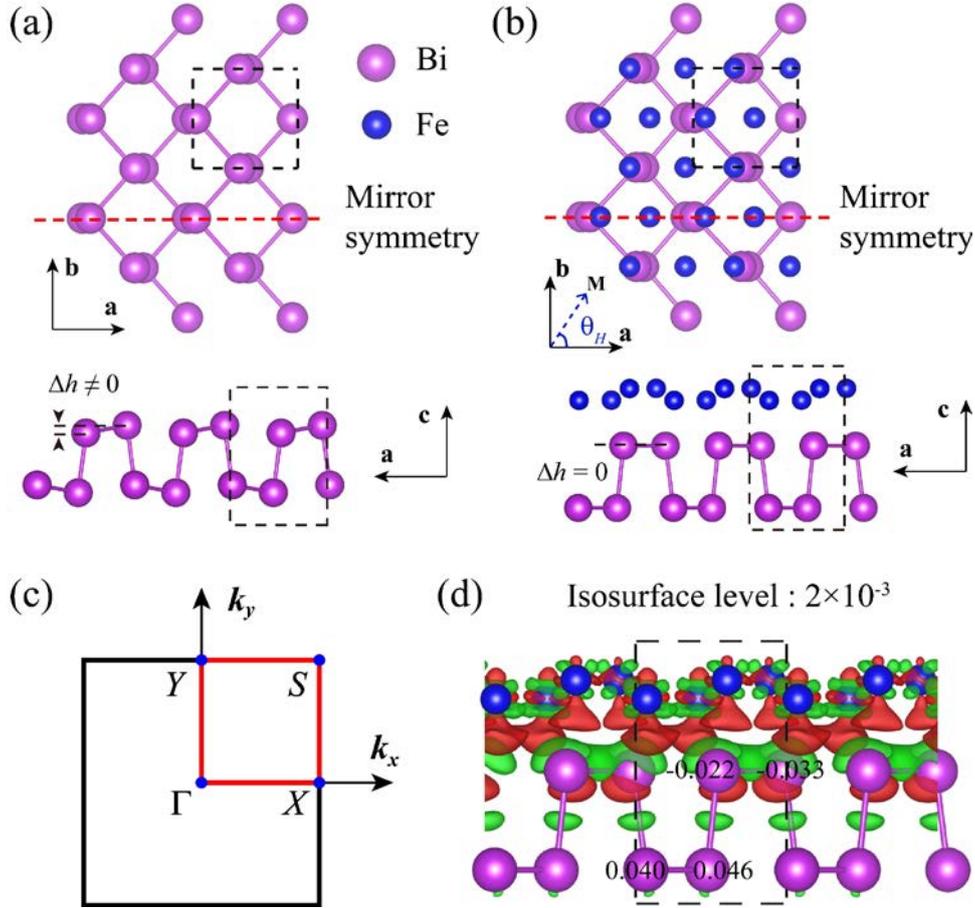

FIG. 1. (a) Crystal structure of BP-Bi ML. Its top and side views are shown in the top and bottom panels, respectively. (b) same as (a) but for the relaxed Bi/1ML-Fe. The inset in (b) shows the definition of the in-plane magnetization orientation. In (a) and (b), the red dashed lines indicate the mirror symmetry. (c) The first Brillouin zone of two-dimensional BP-Bi ML and Bi/Fe heterostructures. (d) Charge density difference $\Delta\rho$ in Bi/1ML-Fe. The red and green show increase and decrease in $\rho$, respectively. The black dashed box shows the unit cell of Bi/1ML-Fe. Numbers give the induced magnetic moments (in units of $\mu_B$) of Bi atoms. In (a)-(c), the black dashed boxes indicate the unit cells of BP-Bi and Bi/1ML-Fe.

To shed some light on the interaction across the Bi-Fe interface, we first examine the charge density difference $\Delta\rho = \rho_{Bi/Fe} - \rho_{Bi} - \rho_{Fe}$. Taking Bi/1ML-Fe as an example, charge transfer between Fe and Bi atoms is noticeable, as shown in Fig. 1(d). Accordingly, Bi atoms have nonzero induced magnetic moments. For the other seven Bi/Fe heterostructures with the thickness of Fe films varying from two to eight MLs, we find that charge transfer takes place between Bi and the interfacial Fe atoms (Fig. S3 in SM). This indicates that there is significant interfacial interaction between BP-Bi ML and Fe film in all Bi/Fe heterostructures. Interestingly, the induced magnetic moments at Bi atoms depend on the thickness of Fe films (Fig. S3 in SM). As shown in Fig. S4 in SM, the averaged magnetic moments of each Fe layer in Bi/Fe heterostructures vary when the thickness of Fe films increases. All of these imply that Gilbert dampings in Bi/Fe heterostructures may change with the thickness of Fe films. In addition, to understand the disappearance of the buckling in Bi/Fe heterostructures, we examine the crystal structure and charge density difference of BP-Bi in contact with inert hexagonal boron nitride (h-BN) [denoted as Bi/h-BN]. We find that Bi/h-BN exhibits minimal charge transfer and maintains small buckling (see Fig. S5 in SM), which contrasts with the sizeable charge transfer and disappearance of buckling in Bi/1ML-Fe. Thus, the disappearance of small buckling of BP-Bi ML in all relaxed Bi/Fe is closely related to the charge transfer at the Bi-Fe interface. This is consistent with previous studies showing that the degree of buckling in two-dimensional materials can be controlled by charge doping [61-63].

We first investigate the Gilbert damping of Bi/1ML-Fe. As shown in Fig. 2(a), the Gilbert damping initially decreases and then increases as we increase the scattering rate $\Gamma$. This trend is consistent with previous studies and can be understood according to the breathing Fermi surface model [20,22,24]. Because Gilbert damping has interband and intraband contributions and the former increases whereas the latter decreases when the scattering rate $\Gamma$ increases, it is a non-monotonic function of the scattering rate $\Gamma$ [15]. Compared with the 1ML-Fe, the Gilbert damping of Bi/1ML-Fe is strongly enhanced by about two orders of magnitude. Hence, the presence of BP-Bi has a remarkable effect on the Gilbert damping of Bi/1ML-Fe. This is perceivable because such significant enhancement in Gilbert damping is mainly due to the strong SOC of BP-Bi, similar to what was found in the previous study for Fe/Bi$_2$Se$_3$ [20].

By setting its in-plane magnetizations along the *a* (i.e., $\theta_H = 0°$) and *b* (i.e., $\theta_H = 90°$) axes, we find that Bi/1ML-Fe exhibits different Gilbert dampings for these two magnetization orientations [Fig. 1(a)]. Such difference clearly indicates that Bi/1ML-Fe has orientationally anisotropic Gilbert dampings. This anisotropy is more obvious at a low scattering rate $\Gamma$, resulting from the smaller smearing for anisotropic interband and intraband contributions [15]. As expected, the orientational anisotropy of Gilbert damping decreases for large $\Gamma$ because of the broadening effect [15]. Considering the practical applications in spintronic devices and the high Curie temperature of Fe films [64,65], we focus on discussing the Gilbert damping of Bi/Fe heterostructures with $\Gamma$ = 26 meV (corresponding to room temperature) hereafter.

To obtain a deeper insight into the anisotropic Gilbert damping of Bi/1ML-Fe, we study its dependence on the in-plane magnetization orientation (i.e., $\theta_H$) and show the results in Fig. 2(b). First of all, the Gilbert damping of Bi/1ML-Fe displays a twofold symmetry with respect to the *a* axis, consistent with its mirror symmetry [Fig. 1(b)]. In addition, the Gilbert damping of Bi/1ML-Fe is invariant when its magnetization is reversed [see Fig. 2(b)]. As a result of such invariance and the mirror symmetry, the Gilbert damping of Bi/1ML-Fe has the twofold symmetry with respect to the *b* axis as well. Finally, it is worth mentioning that Bi/1ML-Fe has maximal and minimal Gilbert dampings when its magnetization is along the *a* and *b* axes, respectively.

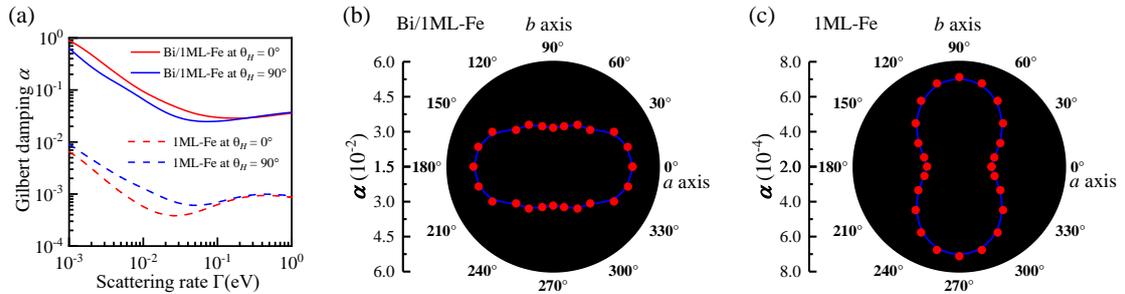

FIG. 2. (a) The $\Gamma$-dependent Gilbert dampings of Bi/1ML-Fe and 1ML-Fe with $\theta_H = 0°$ and $\theta_H = 90°$. (b) The $\theta_H$ dependence of Gilbert dampings of Bi/1ML-Fe at $\Gamma$ = 26 meV. (c) same as (b) but for 1ML-Fe. In (b) and (c), red circles are calculated results and blue lines are guidance for eyes.

Following previous studies [41], we define a maximum-minimum Gilbert damping ratio, $\eta$, to characterize the Gilbert damping anisotropy. According to this definition, we obtain that the $\eta$ of Bi/1ML-Fe is about 154%, which is smaller than that ($\eta = \sim186\%$) of the 1ML-Fe. By comparing the $\theta_H$ dependence of Gilbert damping of Bi/1ML-Fe and the 1ML-Fe [Fig. 2(b)-2(c)], we find that the presence of BP-Bi ML has a remarkable effect on the anisotropy of Gilbert dampings. In Bi/1ML-Fe, the maximal Gilbert damping appears at $\theta_H = 0°$ and the minimal one at $\theta_H = 90°$. In contrast, the Gilbert damping anisotropy of the 1ML-Fe is opposite, namely, maximizing at $\theta_H = 90°$ and minimizing at $\theta_H = 0°$. Taking such change together with the aforementioned enhanced Gilbert dampings, we infer that the SOC of BP-Bi ML plays an important role in determining the Gilbert damping of the Bi/1ML-Fe heterostructure.

In previous studies of CoFe and Heusler alloy Co-Fe-Al films, it is unraveled that there is a close correlation between the intrinsic SOC anisotropy and Gilbert damping anisotropy [41-43]. Hence, the Gilbert damping anisotropy in the 1ML-Fe originates from its intrinsic SOC anisotropy. On the other hand, it was reported that the interfacial SOC strongly depends on the in-plane magnetization orientation of the Fe layer in Fe/GaAs [66]. Moreover, several theoretical and experimental works suggest that interfacial SOC plays a significant role in generating anisotropic Gilbert damping [40,42,45]. Bi/1ML-Fe has a Bi-Fe interface and its maximal Gilbert damping occurs at $\theta_H = 0°$, rather than $\theta_H = 90°$ for the 1ML-Fe. Obviously, the strong interfacial SOC from Bi atoms may overwhelm the intrinsic anisotropic SOC of Fe and alter the Gilbert damping anisotropy in Bi/1ML-Fe. It is worth noting that Rashba SOC does not contribute to the dependence of the Gilbert damping on the in-plane magnetization orientation in Bi/1ML-Fe, since the Rashba SOC does not break spin rotation symmetry within the plane and Bi/1ML-Fe has a weak Rashba SOC with a relatively small Rashba coefficient of 69.9 meV·Å (Fig. S6 in SM [67]).

To understand the mechanism of SOC which alters the anisotropic Gilbert damping in Bi/1ML-Fe, we investigate its band structures and $k$-dependent contributions to its Gilbert damping at $\theta_H = 0°$ and $\theta_H = 90°$. As shown in Figs. 3(a)-3(b), the band structures have obvious differences when its in-plane magnetization orientation changes. Especially, the bands intersect the Fermi levels at different $k$ points and the hybridization between Fe and Bi orbitals at these $k$ points is also different for $\theta_H = 0°$ and $\theta_H = 90°$. By examining the *k*-dependent contributions to the Gilbert damping [Figs. 3(c)-3(d)], we see large contributions from *k* points where bands cross or are near the Fermi level (indicated by arrows in Fig. 3). More explicitly, the dominant contribution to Gilbert dampings is from the $\Gamma - X$ path for $\theta_H = 0°$ whereas it from the $Y - \Gamma$ path for $\theta_H = 90°$. As a result of such dependence of the contributions, Bi/1ML-Fe naturally has different Gilbert dampings for $\theta_H = 0°$ and $\theta_H = 90°$.

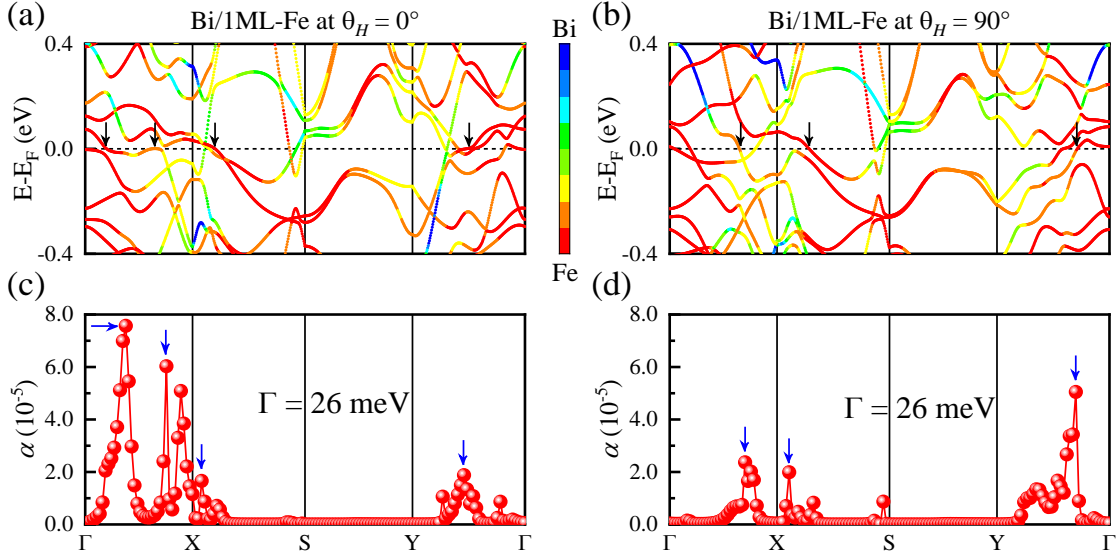

FIG. 3. (a) DFT+SOC calculated band structure and (c) *k*-dependent contributions to the Gilbert damping of Bi/1ML-Fe at $\theta_H = 0°$. (b) and (d) same as (a) and (c) but for $\theta_H = 90°$. In (a) and (b), the Fermi energy levels are indicated by horizontal black dashed lines and the color bars indicate the mixing weights of Fe and Bi atoms.

In fact, based on Eq. (1), it is understandable that only bands near the Fermi level can make major contributions to Gilbert damping due to the presence of two delta functions, $\delta(E_F - E_i)$ and $\delta(E_F - E_j)$. Moreover, Gilbert damping also depends on the matrix element $\langle \psi_i | \frac{\partial \mathbf{H}}{\partial u_\mu} | \psi_j \rangle \langle \psi_j | \frac{\partial \mathbf{H}}{\partial u_\nu} | \psi_i \rangle$ which is closely related to SOC [20]. It is also worth noting that SOC matrix elements between orbitals with different magnetic quantum numbers depend on the orientation of magnetizations [see Eq. (5) in Ref. 20]. Consequently, although 1ML-Fe exhibits similar band structures when it has different magnetization orientations, its bands near the Fermi level originate from different 3*d* orbitals [Fig. S7(a)-7(b) in SM] and thus give rise to different contributions to Gilbert damping [Fig. S7(c)-7(d) in SM], due to the 3$d$ orbital dependent SOC. Similarly, the Fe 3$d$ orbital types and weights in the band structures of Bi/1ML-Fe are dependent on magnetization orientations, as shown in Fig. S8 in SM. On the other hand, as shown in Fe/GaAs, interfacial SOC depends on the in-plane magnetization orientation [66]. This means that interfacial SOC is anisotropic. From Figs. 3(a)-3(b) and Figs. S7-S8 in SM, we see that anisotropic interfacial SOC and intrinsic SOC anisotropy manifest as magnetization orientation dependent band structures. Therefore, anisotropic SOC leads to anisotropic band structures and, thereby, is responsible for anisotropic Gilbert damping in Bi/1ML-Fe and 1ML-Fe.

Now, we study the influence of the thickness of Fe films on the Gilbert damping of Bi/Fe heterostructures. As illustrated in Fig. S9 in SM, we find that Bi/Fe heterostructures with the thickness of Fe films varying from two to eight MLs display similar Γ-dependent Gilbert dampings like in Bi/1ML-Fe. Fig. 4(a) shows the Gilbert dampings of all Bi/Fe heterostructures at Γ = 26 meV with $\theta_H$ ranging from 0º to 180º. It is clear that the Gilbert dampings of Bi/Fe heterostructures decrease when the thickness of Fe films increases. This is understandable since the enhanced Gilbert dampings in Bi/Fe heterostructures arise mainly from the interfacial SOC. As the thickness of Fe films increases in Bi/Fe heterostructures, the influence of the interfacial SOC on Fe layers decreases rapidly away from the interface. Accordingly, the enhancement of the Gilbert damping in Bi/Fe heterostructures due to the interfacial SOC will decay, which is consistent with our calculated results.

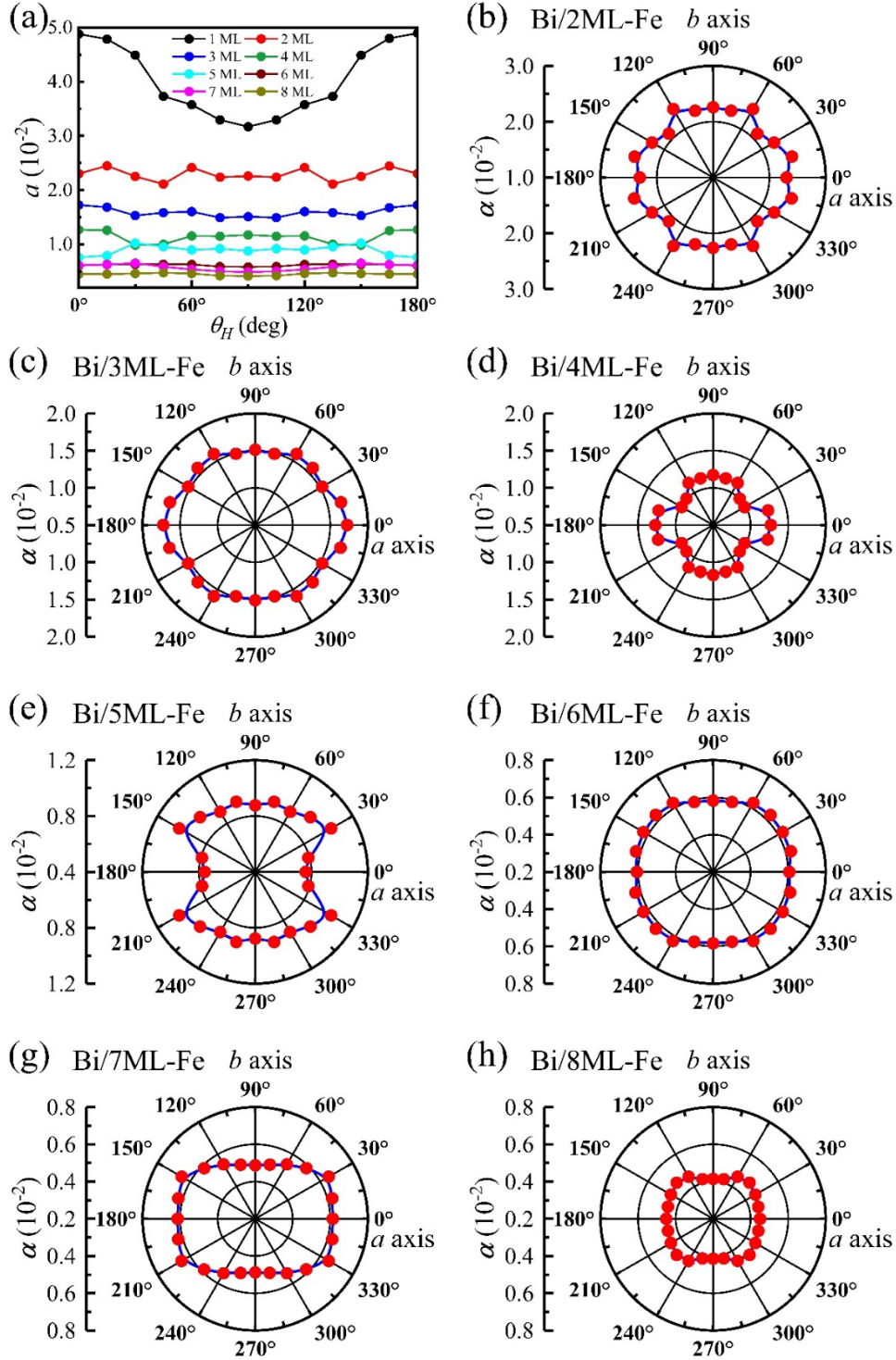

FIG. 4. (a) Fe film thickness-dependent Gilbert dampings of Bi/Fe heterostructures as a function of $\theta_H$ at $\Gamma = 26$ meV. The $\theta_H$ dependence of the Gilbert dampings at $\Gamma = 26$ meV for (b) Bi/2ML-Fe, (c) Bi/3ML-Fe, (d) Bi/4ML-Fe, (e) Bi/5ML-Fe, (f) Bi/6ML-Fe, (g) Bi/7ML-Fe and (h) Bi/8ML-Fe.

To appreciate the effect of the thickness of Fe films on the Gilbert damping anisotropy of Bi/Fe heterostructures, we examine their $\theta_H$-dependent Gilbert dampings and show the results in Figs. 4(b)-4(h). Same as Bi/1ML-Fe, the Gilbert dampings in Bi/Fe heterostructures with the thickness of Fe films varying from two to eight MLs display a twofold symmetry with respect to both the a and b axes. Remarkably, the Gilbert damping anisotropies of Bi/Fe heterostructures have complex relationships with the thickness of Fe films. First of all, the maximal and minimal Gilbert dampings of Bi/Fe heterostructures with different thicknesses of Fe films appear at different angle $\theta_H$. For example, the maximal and minimal Gilbert dampings occur at $\theta_H = 0°$ and $\theta_H = 30°$ in Bi/4ML-Fe [Fig. 4(d)], whereas Bi/5ML-Fe has its maximal and minimal Gilbert dampings at $\theta_H = 30°$ and $\theta_H = 0°$ [Fig. 4(e)]. As the interfacial SOC and the intrinsic SOC anisotropy vary with the thickness of Fe films, the Gilbert damping anisotropy sensitively depends on the thickness of Fe films in Bi/Fe heterostructures. As shown in Fig. 5(a), we see that the damping anisotropy, η, shows a nonmonotonic dependence on the thickness of Fe films. Especially, Bi/1ML-Fe has the strongest Gilbert damping anisotropy. By contrast, Bi/6ML-Fe exhibits a very weak Gilbert damping anisotropy.

Obviously, this trend cannot be understood solely in terms of the weakening of interfacial SOC with increasing the thickness of Fe films. According to our previous discussions [Fig. 2(b)-2(c)], Bi/Fe heterostructures have interfacial SOC from the Bi-Fe interface and the intrinsic SOC from Fe films. Hence, the nearly isotropic Gilbert damping in Bi/6ML-Fe results from the competition between these two factors. To further understand the underlying mechanism, we investigate their band structures and the *k*-dependent contributions to Gilbert dampings [Figs. S10-S16 in SM]. First of all, band structures of Bi/Fe heterostructures with different thicknesses of Fe films exhibit noticeable differences when the in-plane magnetization orientation varies. As found before, the "breathing" of bands around the Fermi level makes a major contribution to Gilbert damping. Therefore, the dependence of band structures on the orientation of magnetization is the cause of anisotropic Gilbert dampings in Bi/Fe heterostructures. As the effect of interfacial SOC on the band structures decays with the thickness of the Fe film, the Gilbert damping anisotropy in Bi/Fe heterostructures also gradually decreases. However, the details of the band structure of Fe films and its intrinsic SOC show some oscillation in the ultrathin regime, and thus the thickness-dependent Gilbert damping anisotropy also oscillates.

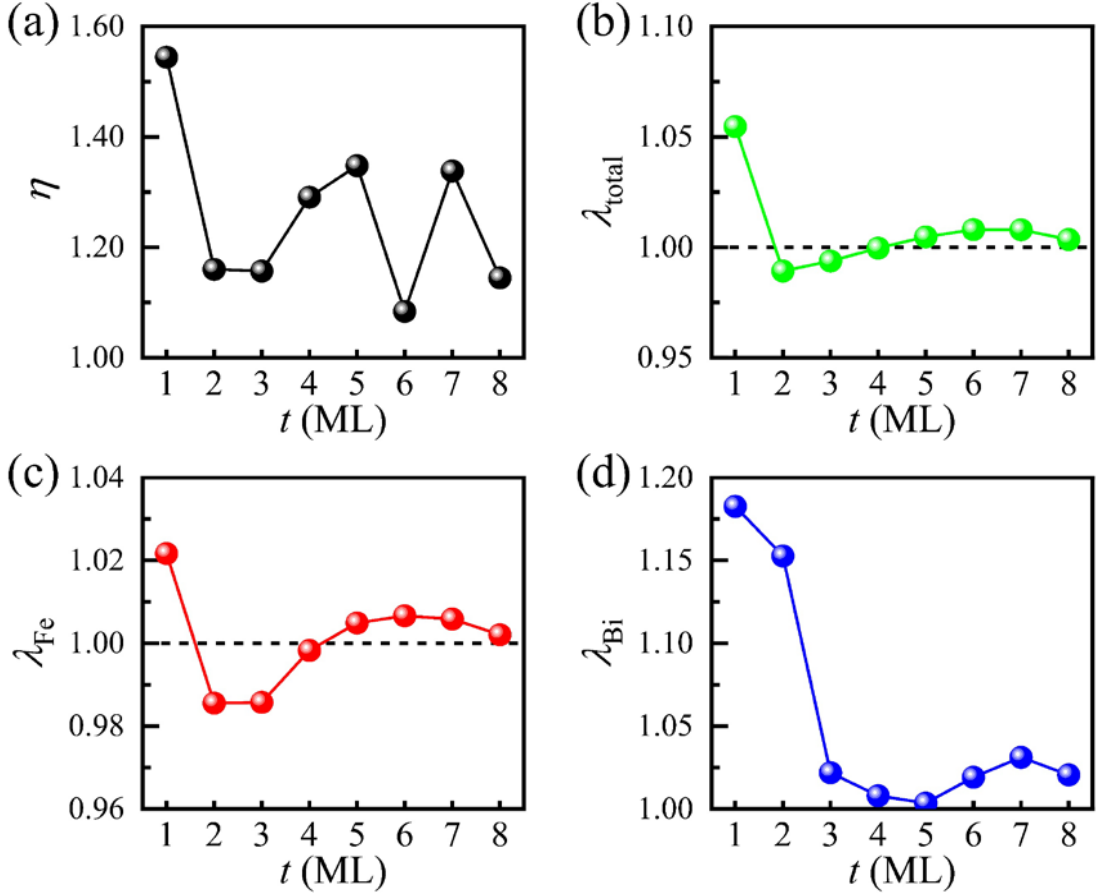

FIG. 5. (a) Gilbert damping anisotropy, $\eta$, as a function of the thickness of Fe films ($t$) in Bi/Fe heterostructures. Fe thickness-dependent anisotropic (b) total, (c) Fe-projected and (d) Bi-projected DOS.

It has been shown in Fe/GaAs that its anisotropic Gilbert damping has a strong correlation with the anisotropic DOSs at the Fermi level [40]. To verify if a similar correlation is applicable to Bi/Fe heterostructures, we calculated their DOSs at the Fermi level for the angles that give the maximal and minimal Gilbert dampings. To feature the anisotropy of DOSs, we define a parameter $\lambda = D_{max}/D_{min}$, where $D_{max}$ ($D_{min}$) is the DOS for the magnetization angle $\theta_H$ with the maximal (minimal) Gilbert damping. As shown in Figs. 5, we observe that the larger anisotropy of DOS reflects the larger Gilbert damping anisotropy in Bi/1ML-Fe. However, for varying thicknesses of Fe films, the anisotropy of DOS is not strongly correlated with the Gilbert damping anisotropy. Especially, for the Bi/2ML-Fe, Bi/3ML-Fe and Bi/4ML-Fe, $D_{max}$ is even less than $D_{min}$ (corresponding to $\lambda < 1$). Considering that the interfacial effect of Bi/Fe heterostructures is strong, it is also meaningful to examine the dependence of Gilbert damping on the anisotropy of their interfacial DOS. Since the interaction between Bi and the first two layers of Fe is relatively strong (see Fig. S3 in SM), we regard the first two layers of Fe as the interface. As shown in Fig. S17, there is no simple correlation between the anisotropy of interfacial DOS and the anisotropy of Gilbert damping. On the whole, the correlation between DOS and Gilbert damping found in Fe/GaAs is not applicable to Bi/Fe heterostructures.

## IV. SUMMARY

In summary, we investigate the thickness-dependent Gilbert dampings of Bi/Fe heterostructures through comprehensive first-principles calculations. Our findings reveal that the presence of BP-Bi ML significantly enhances the Gilbert dampings of Bi/Fe heterostructures by about two orders of magnitude, compared to the unsupported 1ML-Fe. When the thickness of Fe film increases, this enhancement gradually decays, indicating its interfacial nature. Interestingly, the Gilbert damping anisotropy in Bi/Fe heterostructures exhibits a non-monotonic decay with increasing Fe film thickness. We demonstrate that the competition between the strong interfacial SOC and the intrinsic anisotropic SOC of Fe films plays a key role in determining the Gilbert damping anisotropy in these heterostructures. Additionally, we show that these SOC effects manifest as anisotropic band structures, which are the cause of the anisotropic Gilbert dampings. Our work provides profound understandings of anisotropic Gilbert dampings and suggests new possibilities for designing spintronics devices with tailored anisotropic Gilbert damping through ferromagnetic heterostructures.

## ACKNOWLEDGMENTS

This work was supported by the National Key R&D Program of China (Grant No.

2022YFA1403301) and the National Natural Sciences Foundation of China (Grants No. 12104518, 92165204), GBABRF-2022A1515012643. The DFT calculations reported were performed on resources provided by the Guangdong Provincial Key Laboratory of Magnetoelectric Physics and Devices (No. 2022B1212010008) and Tianhe-II. Yusheng Hou acknowledges support from Fundamental Research Funds for the Central Universities, Sun Yat-Sen University (No. 24qnpy108). Ruqian Wu acknowledges support from the USA-DOE, Office of Basic Energy Science (Grant No. DE-FG02-05ER46237).

**References**

[1] T. L. Gilbert, Classics in Magnetics A Phenomenological Theory of Damping in Ferromagnetic Materials, IEEE Trans. Magn. **40**, 3443 (2004).

[2] S. Mangin, D. Ravelosona, J. A. Katine, M. J. Carey, B. D. Terris, and E. E. Fullerton, Current-induced magnetization reversal in nanopillars with perpendicular anisotropy, Nat. Mater. **5**, 210 (2006).

[3] I. M. Miron *et al.*, Fast current-induced domain-wall motion controlled by the Rashba effect, Nat. Mater. **10**, 419 (2011).

[4] A. V. Chumak, V. I. Vasyuchka, A. A. Serga, and B. Hillebrands, Magnon spintronics, Nat. Phys. **11**, 453 (2015).

[5] S. Peng, D. Zhu, J. Zhou, B. Zhang, A. Cao, M. Wang, W. Cai, K. Cao, and W. Zhao, Modulation of Heavy Metal/Ferromagnetic Metal Interface for High-Performance Spintronic Devices, Adv. Electron. Mater. **5**, 1900134 (2019).

[6] O. Ertl, G. Hrkac, D. Suess, M. Kirschner, F. Dorfbauer, J. Fidler, and T. Schrefl, Multiscale micromagnetic simulation of giant magnetoresistance read heads, J. Appl. Phys. **99**, 08S303 (2006).

[7] N. Smith, Micromagnetic modeling of magnoise in magnetoresistive read sensors, J. Magn. Magn. Mater. **321**, 531 (2009).

[8] D. C. Ralph and M. D. Stiles, Spin transfer torques, J. Magn. Magn. Mater. **320**, 1190 (2008).

[9] K. Lee and S. H. Kang, Development of Embedded STT-MRAM for Mobile System-on-Chips, IEEE Trans. Magn. **47**, 131 (2011).

[10] K.-S. Lee, S.-W. Lee, B.-C. Min, and K.-J. Lee, Threshold current for switching of a perpendicular magnetic layer induced by spin Hall effect, Appl. Phys. Lett. **102**, 112410 (2013).

[11] A. Manchon, J. Železný, I. M. Miron, T. Jungwirth, J. Sinova, A. Thiaville, K. Garello, and P. Gambardella, Current-induced spin-orbit torques in ferromagnetic and antiferromagnetic systems, Rev. Mod. Phys. **91**, 035004 (2019).

[12] M. Oogane, T. Wakitani, S. Yakata, R. Yilgin, Y. Ando, A. Sakuma, and T. Miyazaki, Magnetic Damping in Ferromagnetic Thin Films, Jpn. J. Appl. Phys. **45**, 3889 (2006).

[13] I. Garate and A. MacDonald, Gilbert damping in conducting ferromagnets. II. Model tests of the torque-correlation formula, Phys. Rev. B **79**, 064404 (2009).

[14] G. Woltersdorf, M. Kiessling, G. Meyer, J. U. Thiele, and C. H. Back, Damping by slow relaxing rare earth impurities in $Ni_{80}Fe_{20}$, Phys. Rev. Lett. **102**, 257602 (2009).

[15] K. Gilmore, M. D. Stiles, J. Seib, D. Steiauf, and M. Fähnle, Anisotropic damping of the magnetization dynamics in Ni, Co, and Fe, Phys. Rev. B **81**, 174414 (2010).

[16] M. Oogane, T. Kubota, Y. Kota, S. Mizukami, H. Naganuma, A. Sakuma, and Y. Ando, Gilbert magnetic damping constant of epitaxially grown Co-based Heusler alloy thin films, Appl. Phys. Lett. **96**, 252501 (2010).

[17] S. Mizukami *et al.*, Long-lived ultrafast spin precession in manganese alloys films with a large perpendicular magnetic anisotropy, Phys. Rev. Lett. **106**, 117201 (2011).

[18] P. He, X. Ma, J. W. Zhang, H. B. Zhao, G. Lüpke, Z. Shi, and S. M. Zhou, Quadratic Scaling of Intrinsic Gilbert Damping with Spin-Orbital Coupling in $L1_0$ FePdPt Films: Experiments and *Ab Initio* Calculations, Phys. Rev. Lett. **110**, 077203 (2013).

[19] M. A. W. Schoen, D. Thonig, M. L. Schneider, T. J. Silva, H. T. Nembach, O. Eriksson, O. Karis, and J. M. Shaw, Ultra-low magnetic damping of a metallic ferromagnet, Nat. Phys. **12**, 839 (2016).

[20] Y. S. Hou and R. Q. Wu, Strongly Enhanced Gilbert Damping in $3d$ Transition-Metal Ferromagnet Monolayers in Contact with the Topological Insulator $Bi_2Se_3$, Phys. Rev. Appl. **11**, 054032 (2019).

[21] D. Thonig and J. Henk, Gilbert damping tensor within the breathing Fermi surface model: anisotropy and non-locality, New J. Phys. **16**, 013032 (2014).

[22] K. Gilmore, Y. U. Idzerda, and M. D. Stiles, Identification of the dominant precession-damping mechanism in Fe, Co, and Ni by first-principles calculations, Phys. Rev. Lett. **99**, 027204 (2007).

[23] V. Kamberský, Spin-orbital Gilbert damping in common magnetic metals, Phys. Rev. B **76**, 134416 (2007).

[24] L. Qiu, Z. Wang, X.-S. Ni, D.-X. Yao, and Y. Hou, Electrically tunable Gilbert damping in van der Waals heterostructures of two-dimensional ferromagnetic metals and ferroelectrics, Appl. Phys. Lett. **122**, 102402 (2023).

[25] Z. Zhang *et al.*, Strain‐Controlled Spin Wave Excitation and Gilbert Damping in Flexible $Co_2FeSi$ Films Activated by Femtosecond Laser Pulse, Adv. Funct. Mater. **31**, 2007211 (2021).

[26] R. Urban, G. Woltersdorf, and B. Heinrich, Gilbert damping in single and multilayer ultrathin films: role of interfaces in nonlocal spin dynamics, Phys. Rev. Lett. **87**, 217204 (2001).

[27] C. Scheck, L. Cheng, I. Barsukov, Z. Frait, and W. E. Bailey, Low relaxation rate in epitaxial vanadium-doped ultrathin iron films, Phys. Rev. Lett. **98**, 117601 (2007).

[28] H. T. Nembach, J. M. Shaw, C. T. Boone, and T. J. Silva, Mode- and size-dependent Landau-Lifshitz damping in magnetic nanostructures: evidence for nonlocal damping, Phys. Rev. Lett. **110**, 117201 (2013).

[29] T. Kubota, S. Tsunegi, M. Oogane, S. Mizukami, T. Miyazaki, H. Naganuma, and Y. Ando, Half-metallicity and Gilbert damping constant in $Co_2Fe_xMn_{1-x}Si$ Heusler alloys depending on the film composition, Appl. Phys. Lett. **94**, 122504 (2009).

[30] M. Wegscheider, G. Käferböck, C. Gusenbauer, T. Ashraf, R. Koch, and W. Jantsch, Magnetic anisotropy of epitaxial $Fe_{1-x}Si_x$ films on GaAs(001), Phys. Rev. B **84**, 054461 (2011).

[31] Z. Huang *et al.*, Enhancing the Spin-Orbit Coupling in $Fe_3O_4$ Epitaxial Thin Films by Interface Engineering, ACS Appl. Mater. **8**, 27353 (2016).

[32] R. Mandal, Q. Xiang, K. Masuda, Y. Miura, H. Sukegawa, S. Mitani, and Y. K. Takahashi, Spin-Resolved Contribution to Perpendicular Magnetic Anisotropy and Gilbert Damping in Interface-Engineered $Fe/MgAl_2O_4$ Heterostructures, Phys. Rev. Appl. **14**, 064027 (2020).

[33] V. L. Safonov, Tensor form of magnetization damping, J. Appl. Phys. **91**, 8653 (2002).

[34] D. Steiauf and M. Fähnle, Damping of spin dynamics in nanostructures: An *ab initio* study, Phys. Rev. B **72**, 064450 (2005).

[35] M. Fähnle, D. Steiauf, and J. Seib, The Gilbert equation revisited: anisotropic and nonlocal damping of magnetization dynamics, J. Phys. D: Appl. Phys. **41**, 164014 (2008).

[36] J. Seib, D. Steiauf, and M. Fähnle, Linewidth of ferromagnetic resonance for systems with anisotropic damping, Phys. Rev. B **79**, 092418 (2009).

[37] R. Meckenstock, D. Spoddig, Z. Frait, V. Kambersky, and J. Pelzl, Anisotropic Gilbert damping in epitaxial Fe films on InAs(0 0 1), J. Magn. Magn. Mater. **272-276**, 1203 (2004).

[38] R. Yilgin, Y. Sakuraba, M. Oogane, S. Mizukami, Y. Ando, and T. Miyazaki, Anisotropic Intrinsic Damping Constant of Epitaxial $Co_2MnSi$ Heusler Alloy Films, Jpn. J. Appl. Phys. **46**, L205 (2007).

[39] Y. Zhai, C. Ni, Y. Xu, Y. B. Xu, J. Wu, H. X. Lu, and H. R. Zhai, A study on ferromagnetic resonance linewidth of single crystalline ultrathin Fe film grown on GaAs substrate, J. Appl. Phys. **101**, 09D120 (2007).

[40] L. Chen *et al.*, Emergence of anisotropic Gilbert damping in ultrathin Fe layers on GaAs(001), Nat. Phys. **14**, 490 (2018).

[41] Y. Li *et al.*, Giant Anisotropy of Gilbert Damping in Epitaxial CoFe Films, Phys. Rev. Lett. **122**, 117203 (2019).

[42] I. P. Miranda, A. B. Klautau, A. Bergman, D. Thonig, H. M. Petrilli, and O. Eriksson, Mechanisms behind large Gilbert damping anisotropies, Phys. Rev. B **103**, L220405 (2021).

[43] R. Wang *et al.*, Colossal Gilbert Damping Anisotropy in Heusler‑Alloy Thin Films, Adv. Electron. Mater. **9**, 2300049 (2023).

[44] H. Xu, H. Chen, F. Zeng, J. Xu, X. Shen, and Y. Wu, Giant Anisotropic Gilbert Damping in Single-Crystal Co-Fe-B(001) Films, Phys. Rev. Appl. **19**, 024030 (2023).

[45] X. Yang *et al.*, Anisotropic Nonlocal Damping in Ferromagnet/α-GeTe Bilayers Enabled by Splitting Energy Bands, Phys. Rev. Lett. **131**, 186703 (2023).

[46] L. Chen, S. Mankovsky, M. Kronseder, D. Schuh, M. Prager, D. Bougeard, H. Ebert, D. Weiss, and C. H. Back, Interfacial Tuning of Anisotropic Gilbert Damping, Phys. Rev. Lett. **130**, 046704 (2023).

[47] I. Žutić, A. Matos-Abiague, B. Scharf, H. Dery, and K. Belashchenko, Proximitized materials, Mater. Today **22**, 85 (2019).

[48] B. Huang, M. A. McGuire, A. F. May, D. Xiao, P. Jarillo-Herrero, and X. Xu, Emergent phenomena and proximity effects in two-dimensional magnets and heterostructures, Nat. Mater. **19**, 1276 (2020).

[49] J. Gou *et al.*, Two-dimensional ferroelectricity in a single-element bismuth monolayer, Nature **617**, 67 (2023).

[50] G. Kresse and J. Hafner, *Ab initio* molecular dynamics for liquid metals, Phys. Rev. B Condens. Matter **47**, 558 (1993).

[51] G. Kresse and J. Furthmuller, Efficient iterative schemes for *ab initio* total-energy calculations using a plane-wave basis set, Phys. Rev. B Condens. Matter **54**, 11169 (1996).

[52] G. Kresse and J. Furthmüller, Efficiency of ab-initio total energy calculations for metals and semiconductors using a plane-wave basis set, Comput. Mater. Sci **6**, 15 (1996).

[53] J. P. Perdew, K. Burke, and M. Ernzerhof, Generalized Gradient Approximation Made Simple, Phys. Rev. Lett. **77**, 3865 (1996).

[54] P. E. Blochl, Projector augmented-wave method, Phys. Rev. B Condens. Matter **50**, 17953 (1994).

[55] S. Grimme, J. Antony, S. Ehrlich, and H. Krieg, A consistent and accurate *ab initio* parametrization of density functional dispersion correction (DFT-D) for the 94 elements H-Pu, J. Chem. Phys. **132**, 154104 (2010).

[56] A. Brataas, Y. Tserkovnyak, and G. E. Bauer, Scattering Theory of Gilbert damping, Phys. Rev. Lett. **101**, 037207 (2008).

[57] A. Brataas, Y. Tserkovnyak, and G. E. W. Bauer, Magnetization dissipation in ferromagnets from scattering theory, Phys. Rev. B **84**, 054416 (2011).

[58] S. Mankovsky, D. Ködderitzsch, G. Woltersdorf, and H. Ebert, First-principles calculation of the Gilbert damping parameter via the linear response formalism with application to magnetic transition metals and alloys, Phys. Rev. B **87**, 014430 (2013).

[59] X.-P. Wei, X. Zhang, J. Shen, W.-L. Chang, and X. Tao, Gilbert damping, electronic and magnetic properties for quaternary Heusler alloys CrYCoZ: First-principles and Monte Carlo studies, Comput. Mater. Sci **210**, 111453 (2022).

[60] Z. Lu, I. P. Miranda, S. Streib, M. Pereiro, E. Sjöqvist, O. Eriksson, A. Bergman, D. Thonig, and A. Delin, Influence of nonlocal damping on magnon properties of ferromagnets, Phys. Rev. B **108**, 014433 (2023).

[61] Y. Lu *et al.*, Topological Properties Determined by Atomic Buckling in Self-Assembled Ultrathin Bi(110), Nano Lett. **15**, 80 (2014).

[62] C. Xiao, F. Wang, S. A. Yang, Y. Lu, Y. Feng, and S. Zhang, Elemental Ferroelectricity and Antiferroelectricity in Group‑V Monolayer, Adv. Funct. Mater. **28**, 1707383 (2018).

[63] Y. Guo, H. Zhu, and Q. Wang, Large Second Harmonic Generation in Elemental $α$-Sb and $α$-Bi Monolayers, J. Phys. Chem. C **124**, 5506 (2020).

[64] M. Stampanoni, A. Vaterlaus, M. Aeschlimann, and F. Meier, Magnetism of epitaxial bcc iron on Ag(001) observed by spin-polarized photoemission, Phys. Rev. Lett. **59**, 2483 (1987).

[65] W. Durr, M. Taborelli, O. Paul, R. Germar, W. Gudat, D. Pescia, and M. Landolt, Magnetic phase transition in two-dimensional ultrathin Fe films on Au(100), Phys. Rev. Lett. **62**, 206 (1989).

[66] M. Gmitra, A. Matos-Abiague, C. Draxl, and J. Fabian, Magnetic Control of Spin-Orbit Fields: A First-Principles Study of Fe/GaAs Junctions, Phys. Rev. Lett. **111**, 036603 (2013).

[67] H. Yang *et al.*, Significant Dzyaloshinskii–Moriya interaction at graphene–ferromagnet interfaces due to the Rashba effect, Nat. Mater. **17**, 605 (2018).